\documentclass[a4paper]{article}
\usepackage{bigfoot,amsmath,algorithm,listings,authblk}
\usepackage[pdftitle={Formally verified 32- and 64-bit integer division using double-precision floating-point arithmetic},pdfauthor={David Monniaux, Alice Pain}]{hyperref}
\title{Formally verified 32- and 64-bit integer division using double-precision floating-point arithmetic}
\author[1]{David Monniaux}
\author[1,2]{Alice Pain}
\affil[1]{Univ. Grenoble Alpes, CNRS, Grenoble INP, VERIMAG\\38000 Grenoble, France}
\affil[2]{École normale supérieure, Paris}

\begin{document}
\maketitle

\begin{abstract}
  Some recent processors are not equipped with an integer division unit.
  Compilers then implement division by a call to a special function supplied by the processor designers, which implements division by a loop producing one bit of quotient per iteration.
  This hinders compiler optimizations and results in non-constant time computation, which is a problem in some applications.
  
  We advocate instead using the processor's floating-point unit, and propose code that the compiler can easily interleave with other computations.
  We fully proved the correctness of our algorithm, which mixes floating-point and fixed-bitwidth integer computations, using the Coq proof assistant and successfully integrated it into the CompCert formally verified compiler.
\end{abstract}

\section{Introduction}
Some instruction sets (x86, AArch64, RISC-V M extension…) feature integer division instructions. Even if all other operations (integer, floating-point, memory…) are fully pipelined, division is typically handled differently: only one division can be handled at a given time (no pipelining), and the execution time of the division operation depends on the operands.
This makes the processor design more complex, and also precludes constant-time execution, which is desirable in some contexts, for instance for safety-critical control systems%
\footnote{Such systems favor \emph{predictable} architectures.
  Constant-time execution simplifies worst-case execution time (WCET) analysis. Static analysis of WCET typically relies on explicit enumeration of reachable pipeline states, and instructions with operand-dependent execution time increase the number of such states and may lead to combinatorial explosion.}
and in systems where timing attacks are a concern.

In contrast, some architectures eschew divisor units and emulate division in software, totally or partially.
Kalray's KV3 processor, the latest in a series of VLIW (very large instruction word) processors, does not have a full division unit. Instead, it has a fully pipelined, correctly rounded IEEE-754 single-precision (\emph{binary32}) floating-point reciprocal operation.
Using this operation as a starting point, we compute a higher-precision approximation of the reciprocal, then correct 32-bit and 64-bit integer quotients, using the processor's IEEE-754 double-precision (\emph{binary64}) floating-point unit and integer operations.

CompCert%
\footnote{See \url{https://compcert.org/}}
is a formally verified compiler for the C language.%
\footnote{Another frontend, known as Velus, exists for this compiler for a subset of the Lustre / Scade synchronous data-flow language, used to implement control systems in industries such as avionics.}
Here, ``formally verified'' means that there is a proof, in the Coq proof assistant, that the execution of the assembly code generated by the compiler matches that of the source code~\cite{leroy09}.

The backend for the Kalray KV3 processor \cite{DBLP:journals/pacmpl/SixBM20}, when compiling integer division, by default generates calls to special library functions.
These functions are based on code, provided by Kalray, that computes divisions by looping over a special arithmetic instruction that performs one step of division and computes one bit of the quotient.%
\footnote{The comments in the library mention that this approach is adapted from~\cite{TMS320C60000_division}. Also, when CompCert for KV3 determines that the divisor of a 32-bit division is constant, it does not generate this function call and instead produces a sequence of integer operations involving multiplications~\cite{DBLP:conf/pldi/GranlundM94} which is proved to be equivalent to division by this divisor.}
These calls are axiomatized inside CompCert to return the correct quotient and remainder:
CompCert trusts that they perform correctly as intended.
This, arguably, breaks CompCert's design; but anyway CompCert has to trust that the processor actually implements its own instructions correctly; trusting that a simple and understandable integer arithmetic procedure published by the processor designers is correct is not a far stretch from trusting that the processor hardware behaves correctly.%
\footnote{One workaround would be to prove formally correct this snippet of code, whether directly using CompCert's semantics or using an external tool for reasoning on C source code.}

The situation was however different for our new algorithm. During its design, especially for the 64-bit version, we came across a number of corner cases that could have been unnoticed by less careful testing.
Caution dictated to be wary that there could be more corner cases.
It would not have been right to add an axiom that this new algorithm was correct. Instead we resolved to fully prove it correct inside Coq, based on the definitions of the integer and IEEE-754 instructions that we use.
We prove that, when the divisor is nonzero, our sequence of operations returns the correct quotient (respectively, remainder), without introducing any new axiom.

\section{Division Algorithms}
We provide division algorithms for unsigned 32-bit and 64-bit integers.
We assume the processor supports IEEE-754 double-precision (\emph{binary64}) operations and some single-precision (\emph{binary32}) operations:
an IEEE-754 single-precision reciprocal ($x \mapsto 1/x$) instruction,%
\footnote{It seems possible to adapt proofs to less precise approximate reciprocal operations, since we don't use the full precision that we prove.} 
conversions between 64-bit signed and unsigned integers and IEEE-754 double-precision, conversions between single- and double-precision floating-point numbers, double-precision fused multiply-add (fma).
All operations should be rounded to the nearest target value; the rule used to break ties between equally near numbers is unimportant.%
\footnote{Because we had to completely specify the rounding mode for our formal proofs, we picked breaking ties to even numbers, as this is the most common rounding mode. We however do not use this anywhere in the proofs.}
To deal with special cases, we use conditional moves (if-then-else functional statements), maintaining constant-time execution; these may be replaced by normal if-then-else control-flow, if needed.

Signed division following C semantics (quotients truncated towards zero) is implemented by calling unsigned division on the absolute values of the dividend and divisor and adjusting signs afterwards.

\subsection{32-bit Division}
To compute the quotient of $a$ by $b$, we first compute a single-precision reciprocal of the divisor $b$, thus with 23 bits of precision. Then we follow a well-known approach \cite{DBLP:conf/tphol/Harrison00} to obtain approximately 46 bits of precision, using one step of a fixed-point iteration leading to the reciprocal, implemented using two double-precision fused multiply-add operations. 46 bits of precision is more than enough to compute a very precise approximation of the quotient $a/b$ by multiplying with the dividend~$a$.
Because we do not know if that quotient was approximated by above or below, we compute the remainder associated with it using an integer multiply-add, and adjust the quotient if that remainder is negative (Algorithm~\ref{alg:32-bit}).

In the following, \lstinline|fma| is fused multiply-add ($\lstinline|fma|(x, y, z) \simeq xy + z$). We also use special built-in operators for converting double precision numbers to signed and unsigned 64-bit numbers with round to nearest (the method of breaking ties is unimportant).

\begin{algorithm}
\caption{32-bit unsigned division}\label{alg:32-bit}
\begin{lstlisting}[firstline=4]
uint32_t div32(uint32_t a, uint32_t b) {
  float bs = (float) b;
  double bd = (double) b;
  float invbs0 = 1.0f / bs;
  double invbd0 = (double) invbs0;
  double alpha = fma(-bd, invbd0, 1.0);
  double invbd = fma(alpha,invbd0,invbd0);
  double ad = (double) a;
  double qd = ad * invbd;
  // round to nearest, unsigned
  uint64_t q0 = __builtin_luround_ne(qd);
  int64_t r0 = a - b * q0;
  uint64_t ql = r0<0 ? q0-1 : q0;
  return (uint32_t) ql;
}
\end{lstlisting}
\end{algorithm}

If necessary, the remainder can be computed as
\begin{lstlisting}
uint32_t r2 = (uint32_t) r0;
uint32_t r = r0<0 ? r2+b : r2; 
\end{lstlisting}

Remark that most of the expensive computation depends only on~$b$: we postpone multiplication by $a$ until the last moment.
It would be possible to start by computing an approximation of $a/b$ instead of $1/b$, and refine that approximation, but this would not save any operation (we still would need a multiplication by $a$), and this would preclude the compiler from hoisting computations sharing the same divisor.

\subsection{64-bit Division}
The natural generalization of the 32-bit algorithm would be to increase the number of fixed-point iterations to compute a more precise approximation of the reciprocal of the divisor, but, since IEEE-754 double-precision only has 53 bits of significand, this would be insufficient to compute correct quotients of 64-bit numbers.
Instead, Algorithm~\ref{alg:64-bit} proceeds in three steps:
\begin{enumerate}
\item compute an approximation $q_1$ of $a/b$, and associated remainder $r_1=a-bq_1$ using the single-precision reciprocal;
\item compute an approximation $q_2$ of $r_1/b$, and associated remainder $r_2=r_1-bq_2$, using the more precise reciprocal approximation from the preceding subsection;
\item adjust the quotient if $r_2$ is negative.
\end{enumerate}
Then $q_0=q_1+q_2$ is the quotient $q=\lfloor a/b\rfloor$ in most cases (if $2 \leq b < 2^{63}$).
Note that the precise reciprocal approximation can be computed in parallel to~$r_1$.

The cases $b=1$ (return $q=a$) and $b \geq 2^{63}$ (return $q=1$ if $a \geq b$, $q=0$ otherwise) are treated separately;
\begin{itemize}
\item $b=1$ means that $r_1 \simeq a$ (not necessarily equal, since large values of $a$ would incur rounding when converted to floating-point), and $r_1$ would not fit within a signed 64-bit integer; in this case we directly output $q=a$;
\item if $b \geq 2^{63}$ and $q_1=1$, $r_1$ may not fit within a 64-bit signed integer; e.g., $a = 2^{63}$ and $b=2^{64}-1$;
  since in this case $a < 2b$, the quotient is $0$ or $1$ depending on whether $a < b$.
\end{itemize}
These special cases are treated in parallel to the main computation, and the special result is substituted, if applicable, using a 1-cycle conditional move at the end.

Note that $b \geq 2^{63}$ if and only if it is negative if considered as a 64-bit signed number, and that $q=1$ if $a \geq b$, $q=0$ otherwise amounts to taking $a \geq b$ as a truth value. This may help simplify the assembly code.

\begin{algorithm}
\caption{64-bit unsigned division}\label{alg:64-bit}
\begin{lstlisting}[firstline=5]
uint64_t div64(uint64_t a, uint64_t b) {
  double bd = (double) b;
  float bs = (float) bd;
  float invbs0 = 1.0f / bs;
  double invbd0 = (double) invbs0;
  double alpha = fma(-bd, invbd0, 1.0);
  double invbd = fma(alpha,invbd0,invbd0);
  double ad = (double) a;
  double q1d = ad * invbd0;
  // round to nearest, unsigned
  uint64_t q1 = __builtin_luround_ne(q1d);
  int64_t r1 = a - b*q1;
  double r1d = (double) r1;
  double q3d = r1d * invbd;
  // round to nearest, signed
  int64_t q3 = __builtin_lround_ne(q3d);
  int64_t r3 = r1 - b * q3;
  int64_t q2 = r3<0 ? q3-1 : q3;
  uint64_t q0 = q1 + q2;
  bool is_big = (int64_t) b < 0; //b>=2^63
  uint64_t if_big = a >= b;
  bool is_one = b <= 1;
  uint64_t special = is_big ? if_big : a;
  return (is_one||is_big) ? special : q0;
}
\end{lstlisting}
\end{algorithm}

Here, we assume that conversion from floating-point numbers to integers do not trap (do not produce an exception stopping the program) if the number does not fall within the target range, and instead returns an ``undefined'' value.%
\footnote{Here, we just assume the ``undefined'' value results in further ``undefined'' values through further computations. If the ``undefined'' value can be assumed to be a valid 64-bit integer, then we may simplify our code a tiny bit.}
If this operation may trap, it is necessary to rewrite the functional if-then-else (conditional move) into a control-flow test (Algorithm~\ref{alg:64-bit-tests}), which breaks constant-time execution.

\begin{algorithm}
\caption{64-bit unsigned division avoiding trapping conversions by branching out}\label{alg:64-bit-tests}
\begin{lstlisting}[firstline=5]
uint64_t div64(uint64_t a, uint64_t b) {
  if (b <= 1) return a;
  if ((int64_t) b < 0) // b >= 2^63
    return a >= b;
  double bd = (double) b;
  float bs = (float) bd;
  float invbs0 = 1.0f / bs;
  double invbd0 = (double) invbs0;
  double alpha = fma(-bd, invbd0, 1.0);
  double invbd = fma(alpha,invbd0,invbd0);
  double ad = (double) a;
  double q1d = ad * invbd0;
  // round to nearest, unsigned
  uint64_t q1 = __builtin_luround_ne(q1d);
  int64_t r1 = a - b*q1;
  double r1d = (double) r1;
  double q3d = r1d * invbd;
  // round to nearest, signed
  int64_t q3 = __builtin_lround_ne(q3d);
  int64_t r3 = r1 - b * q3;
  int64_t q2 = r3<0 ? q3-1 : q3;
  return q1 + q2;
}
\end{lstlisting}
\end{algorithm}

\section{Proof of Correctness}
Our proofs rely on the formalization of IEEE-754 in the Flocq library%
\footnote{\url{https://flocq.gitlabpages.inria.fr/}}~%
\cite{DBLP:conf/arith/BoldoM10,DBLP:books/daglib/0041425,HDR_Boldo,Boldo_Melquiond_ARITH21}.
In particular, Flocq has
\begin{itemize}
\item executable definitions of individual IEEE-754 operations, which are used to specify CompCert's C and assembler floating-point semantics (these definitions compute the bit pattern in the output as a function of the inputs)
\item proofs that these definitions match the non-executable%
\footnote{This specification is not executable in the sense that it involves Coq's real numbers. Since in this article we are only concerned about addition, subtraction, multiplication and division, it could be possible to write this specification using only rational numbers, which would make it executable.}
  specification that an IEEE-754 operation amounts to computing the operation in the reals then rounding into the appropriate type.
\end{itemize}

For most architectures, CompCert does very little proofs about floating-point: an operation having a certain semantics in the source language (say, single-precision floating-point addition) is translated into an operation with the same semantics in the assembly language. There are proofs about how to implement certain conversion operations using simpler operations for architectures that do not support these directly as individual instructions. There are however no proofs about replacing operations by combinations of other operations involving fine points about roundoff error, as we need here.

To reduce the proof effort on bounding roundoff error, we heavily rely on the Coq \verb|gappa| tactic, which calls the Gappa tool%
\footnote{\url{https://gappa.gitlabpages.inria.fr/}}~%
\cite{DBLP:journals/toms/DaumasM10,DBLP:books/daglib/0027236,PhD_Melquiond,HDR_Melquiond,Boldo_Melquiond_ARITH21}.
We also heavily use Coq's \verb|ring| and \verb|field| arithmetic equality tactics and the \verb|lia| linear integer arithmetic tactic for inequalities.

\subsection{Approximate Double-Precision Reciprocal}
The computation of the approximate reciprocal that we use in our algorithms is often presented as a case of Newton-Raphson iteration. Let us give a different intuition here.
In order to iteratively refine a numerical solution, we express the result we want, $1/b$, as a fixed-point of a contracting function $f$ chosen such that $f(1/b)=1/b$.
The simplest kind of contracting function (approximated by fma) is $f(x)=\alpha x + \beta$ with small~$\alpha$.
What $\alpha$ to pick? Assume $\tilde{b}_0 \simeq 1/b$, then $\alpha=1-b\tilde{b}_0 \simeq 0$ (also approximable by fma) is small. Solve $f(1/b)=1/b$ for $\beta$: $\beta=\tilde{b}_0$.
We then have $|f(x)-1/b| \leq \alpha|x-1/b|$.
$\tilde{b}=f(\tilde{b}_0)$ is thus an ever better approximation of~$1/b$ than~$\tilde{b}_0$.
Let us now see how to turn this reasoning over real numbers into a result on floating-point computations.

We prove in theorem \verb|approx_inv_longu_correct_rel| that \verb|invbd| is a very precise approximation of the reciprocal of \verb|b|, with relative error $\frac{\texttt{invbd}-1/b}{1/b}$ less than $1049\times2^{-56} \simeq 2^{-46}$.

Let us comment the proof approach:
when we encounter an expression $r_d(e)$ (respectively, $r_s(e)$), meaning ``the double-precision rounding of $e$'', we replace it by $e (1 + \epsilon_e)$ and we use the \verb|gappa| Coq tactic to bound the relative error $\epsilon_e$;
we simplify expressions in the real field using Coq's \verb|field| tactic.
In the end, the final relative error is expressed as a polynomial in the relative errors of the individual operations, and easily bounded.

From this relative error, we prove that $q_d$ in the 32-bit algorithm is such that $|q_d-a/b| < 1/2$, whence the correct quotient after adjusting for negative remainder~$r_0$.

\subsection{64-bit Division Algorithm}
The proof of the 64-bit division algorithm is more involved, and distinguishes four cases: $b=1$, $2 \leq b \leq 2^{42}$, $2^{42} < b < 2^{63}$, and $b \geq 2^{63}$.
The first and last cases are dealt with by explicit tests in the algorithm, respectively by answering $q=a$ and $q=(\text{if~} a\geq b \text{~then~} 1 \text{~else~} 0)$.

The proof of the remaining cases is more complex than for the 32-bit algorithm. One reason is that, contrary to what happens with 32-bit operands, large values of $a$ and $b$ cannot be in general represented exactly within double-precision arithmetic, so we have to deal with the roundoff error induced by the conversions of~$a$ and $b$ in addition to the roundoff error induced by the later operations.

\paragraph{Small $b$: $2 \leq b \leq 2^{42}$}
We prove that $|r_1| \leq 44\times 10^{11}$.
We then prove that, if $|r_1| \leq 342\times 10^{11}$, then $q_2$ truly is the quotient of the division of $r_1$ by~$b$:
because the numerator $r_1$ has small magnitude, the resulting quotient $q_{3d}$ also has small magnitude and the relative error on $q_{3d}$ translates into an absolute error less than $1/2$.
The result follows.

\paragraph{Big $b$: $2^{42} < b < 2^{63}$}
We prove that $q_2$ truly is the quotient of the division of $r_1$ by~$b$:
because the denominator $b$ is large, the resulting quotient $q_{3d}$ has small magnitude and the relative error on $q_{3d}$ translates into an absolute error less than $1/2$.
The result follows.
Note that we do not prove anything about~$r_1$ in this context.

\subsection{Correspondence with IEEE-754 Numbers}
So far, we have expressed floating-point computations as compositions of operations of real numbers and rounding operators.
This ignores the fact that IEEE-754 floating-point values may be infinite, or ``not a number'' (NaN), which is the case of the IEEE-754 datatypes as used in CompCert's semantics.
For each operation on IEEE-754 numbers in CompCert, we invoke a correctness theorem in Flocq (e.g., \verb|Bfma_correct| for fma) that says that if the operands are finite (meaning, neither infinities nor NaN) and the result of the operation over the reals fits the maximal magnitude accepted by the target type, then the result of the operation is finite and has a real value, which is the correctly rounded value of the result of the operation applied to the operands.
In order to apply these theorems, one thus has to prove lemmas on the magnitudes of the computed numbers.
Most of the size of our proofs results from auxiliary lemmas on magnitudes of numbers.

\section{Performance Evaluation}
To compare performance with the previous method proposed by Kalray,
we computed 64-bit quotients with $a = 2^{40} + 222823k$
and $b = 2^{12} + 19k$, for $0 \leq k < 10000$
and timed the time to compute these $10000$ quotients.
In the following, numbers are clock cycles (less is faster), \emph{Loop} is an implementation of division using a hardware loop over Kalray's special instruction producing one bit of quotient, \emph{Floating-point} is our floating-point algorithm, and we consider loop structures that both compute the same $10000$ quotients, one with one quotient per iteration, the other with two quotients per iteration (thus $5000$ iterations).

\begin{center}
\begin{tabular}{lrr}
  Method          & Loop & Floating-point \\
  \hline
  One quotient per iteration & 620180 & 522316\\
  Two quotients per iteration & 589696 & 292314\\
\end{tabular}
\end{center}

When two quotients are computed for each loop iteration, CompCert can schedule \cite{DBLP:journals/pacmpl/SixBM20} together the instructions that compute the two quotients, thus the nearly halved computation time.
Recall that the processor is fully pipelined, meaning that if a floating-point instruction enters the pipeline, another floating-point instruction may enter the pipeline at the next clock cycle even though the previous instruction has not yet produced a result, as long as the second instruction does not depend (from its operands) on the first instruction.
Several independent computations can thus be weaved together by the compiler, as long as they do not use control-flow (loops and if-then-else).
In contrast, two calls to the function implementing division using a loop cannot be scheduled together.%
\footnote{What would be needed is to inline the called function, then fuse together the two loops in one single loop, which is difficult given that they have different interaction counts.}

For 32-bit quotients, we used $a = 2^{24}+871k$ and $b = 2^{12} + 19k$, for $0 \leq k < 10000$.

\begin{center}
\begin{tabular}{lrr}
  Method          & Loop & Floating-point \\
  \hline
  One quotient per iteration & 469969 & 442101 \\
  Two quotients per iteration & 434501 & 232124 \\
\end{tabular}
\end{center}

If the same divisor is used for all iterations, the loop-invariant code motion optimization presented in \cite{10.1145/3529507} can move all computations involving the divisor only out of the loop.
Here are cycles counts when $b=74567$, for 64-bit operands:

\begin{center}
\begin{tabular}{lrr}
  Method          & Loop & Floating-point \\
  \hline
  One quotient per iteration & 608158  & 342951 \\
  Two quotients per iteration & 582948 & 237857 \\
\end{tabular}
\end{center}

And for 32-bit operands:%
\footnote{If the divisor is constant and known at compile-time, CompCert replaces 32-bit integer division by a specialized sequence of purely integer operations~\cite{DBLP:conf/pldi/GranlundM94}. We arranged for these benchmarks that it should not be the case.}

\begin{center}
\begin{tabular}{lrr}
  Method          & Loop & Floating-point \\
  \hline
  One quotient per iteration & 458100 & 213014 \\
  Two quotients per iteration & 433000 & 112906 \\
\end{tabular}
\end{center}

\section{Related Work}
The possibility of iterative refinement of reciprocals, quotients and square roots by Newton-Raphson iterations implemented by fused multiply-add has long been recognized~\cite{DBLP:conf/arith/Cornea-HaseganGM99}\cite[ch.~5]{DBLP:books/daglib/0027236}.

The IA-64 architecture did not have division instructions, and much work was done on efficient floating-point and integer division algorithms for this architecture and associated formal proofs~\cite{Divide_IA64,DBLP:conf/tphol/Harrison00,Cornea_et_al_RNC4}.
These algorithms are generally not applicable to the KV3 and other current architectures, since they assume the availability of 82-bit extended precision floating-point operations with 64-bit significands.

\section{Conclusion and Perspectives}
We have successfully formally verified 32-bit and 64-bit integer division algorithms for the Kalray KV3 processor. The algorithms are applicable to any processor with double-precision floating-point arithmetic featuring a fused multiply-add, using round-to-nearest.
The algorithms were implemented in a version of the CompCert verified compiler for the~KV3 available online.%
\footnote{\url{https://gricad-gitlab.univ-grenoble-alpes.fr/certicompil/compcert-kvx.git}, commit \verb|d5f60d87|. The proofs are in files \verb|kvx/FPDivision32.v| and \verb|kvx/FPDivision64.v|. The new division operators are accessible from C using the builtins
  \lstinline|__builtin_fp_udiv32|,
  \lstinline|__builtin_fp_udiv64|,
  \lstinline|__builtin_fp_umod32|
  \lstinline|__builtin_fp_umod64|,
  \lstinline|__builtin_fp_sdiv32|,
  \lstinline|__builtin_fp_sdiv64|,
  \lstinline|__builtin_fp_smod32|,
  \lstinline|__builtin_fp_smod64|.
Since the performance of these new operators is very satisfying, we will use them by default in future releases. External calls to the loop function may be reinstated by the options \verb|-fdiv-i32= stsud| and \verb|-fdiv-i64= stsud|.}
The implementation and proofs take up 883 lines for the 32-bit division, 2670 for the 64-bit division. This size could probably be reduced through refactoring and custom proof automation.
For each bit width, we cover signed and unsigned division and modulo.
In each case, the final theorem states that, for all inputs, our sequence of operations (at the level of the compiler's intermediate representation; these operations map one-to-one to assembly instructions) computes exactly the same value as the corresponding C division or modulo operation when the divisor is nonzero.

Experiments show that our 32-bit and 64-bit constant-time divisions are on average faster than the special functions previously provided by Kalray.
In addition, since our computation is straight-line, as opposed to the loop inside that special function, it can be interleaved with other computations (including other divisions) by the compiler's instruction scheduler. 
Since most of the computation depends only on the divisor, common subexpression elimination by the compiler will simplify computations if several divisions use the same divisor;
similarly, the code depending only on the divisor may be hoisted out of a loop if the divisor remains constant across iterations.

Currently, Kalray's compilers implement floating-point division through calls to \verb|libgcc|'s software floating-point routines, which are themselves implementing by integer arithmetic. A natural extension of our work would be to design, implement and prove correct algorithms using the hardware floating-point unit and especially its fused multiply-add instruction, as it was done for the~IA-64.

\section*{Acknowledgments}
We wish to thank Cyril Six for help in running experiments on actual KV3 processors.

\bibliographystyle{plain}
\bibliography{FP_int_division}
\end{document}